\documentclass[twocolumn]{article}
\usepackage{layout}
\usepackage{moreverb,url}
\usepackage[utf8]{inputenc}
\usepackage{pgfplots}
\usepackage{etoolbox}
\usepackage{caption}
\usepackage{subcaption}
\usepackage{graphicx} 
\usepackage{hyphenat}
\usepackage{booktabs}
\usepackage{balance}
\usepackage{float}
\usepackage{verbatim} 
\usepackage[export]{adjustbox}
\usepackage{array}
\usepgfplotslibrary{dateplot}
\usepackage{filecontents}
\usepackage{lipsum}
\usepackage{mathtools}
\usepackage[colorinlistoftodos]{todonotes}
\usepackage{amsmath}
\usepackage{amsmath}
\pgfplotsset{compat=1.13}
\usepackage{amsmath}
\usepackage{tikz}
\usetikzlibrary{svg.path}
\usepackage{hyperref}
\usepackage{abstract}
\usepackage[affil-it]{authblk}
\date{}
\usepackage[square,numbers]{natbib}
\begin{document}
\title{Characterizing Long-Running Political Phenomena on Social Media} 
\author{Emre Calisir\qquad Marco Brambilla \\ \small \textnormal{Dipartimento di Elettronica, Informazione e Bioingegneria}\\\textnormal{Politecnico di Milano, Milan, Italy} \\
\textnormal{Email:\{firstname.lastname\}@polimi.it}}
\twocolumn[
  \begin{@twocolumnfalse}
    \maketitle
    \begin{abstract}
    Social media provides many opportunities to monitor and evaluate political phenomena such as referendums and elections. In this study, we propose a set of approaches to analyze long\hyp{}running political events on social media with a real\hyp{}world experiment: the debate about Brexit, i.e., the process through which the United Kingdom activated the option of leaving the European Union. We address the following research questions: Could Twitter-based stance classification be used to demonstrate public stance with respect to political events? What is the most efficient and comprehensive approach to measuring the impact of politicians on social media? Which of the polarized sides of the debate is more responsive to politician messages and the main issues of the Brexit process? What is the share of bot accounts in the Brexit discussion and which side are they for? By combining the user stance classification, topic discovery, sentiment analysis, and bot detection, we show that it is possible to obtain useful insights about political phenomena from social media data. We are able to detect relevant topics in the discussions, such as the demand for a new referendum, and to understand the position of social media users with respect to the different topics in the debate. Our comparative and temporal analysis of political accounts can detect the critical periods of the Brexit process and the impact they have on the debate.
\\
\newline


\end{abstract}
\hspace{8pt}

  \end{@twocolumnfalse}
  ]
  
\section{Introduction}
Social media provides many opportunities to monitor and evaluate political phenomena such as referendums and elections. Citizens from all around the world, voters, politicians, private and public authorities participate and contribute to debates on social media platforms with tremendous interest. According to a survey, 66\% of social media users have employed these platforms to post their thoughts about civic and political issues, react to others' postings, press friends to act on issues and vote, follow candidates, like and link to others' content, and belong to groups formed on social networking sites \cite{Rainie-2012}. In this context, Twitter is known as one of the most convenient social media platforms with its prominent features including hashtag based information annotation and retrieval, mention\hyp{}based people referring and re-tweet/like based agreement on the opinions. Segesten and Bossetta found that that citizens \hyp{} not political parties \hyp{} are the primary initiators and sharers of political calls for action leading up to the 2015 British General Elections \cite{Segesten-2016}.

\begin{figure}[t]
\centering
\includegraphics[clip, trim=0cm 0.5cm 0cm 0cm, width=0.5\textwidth]{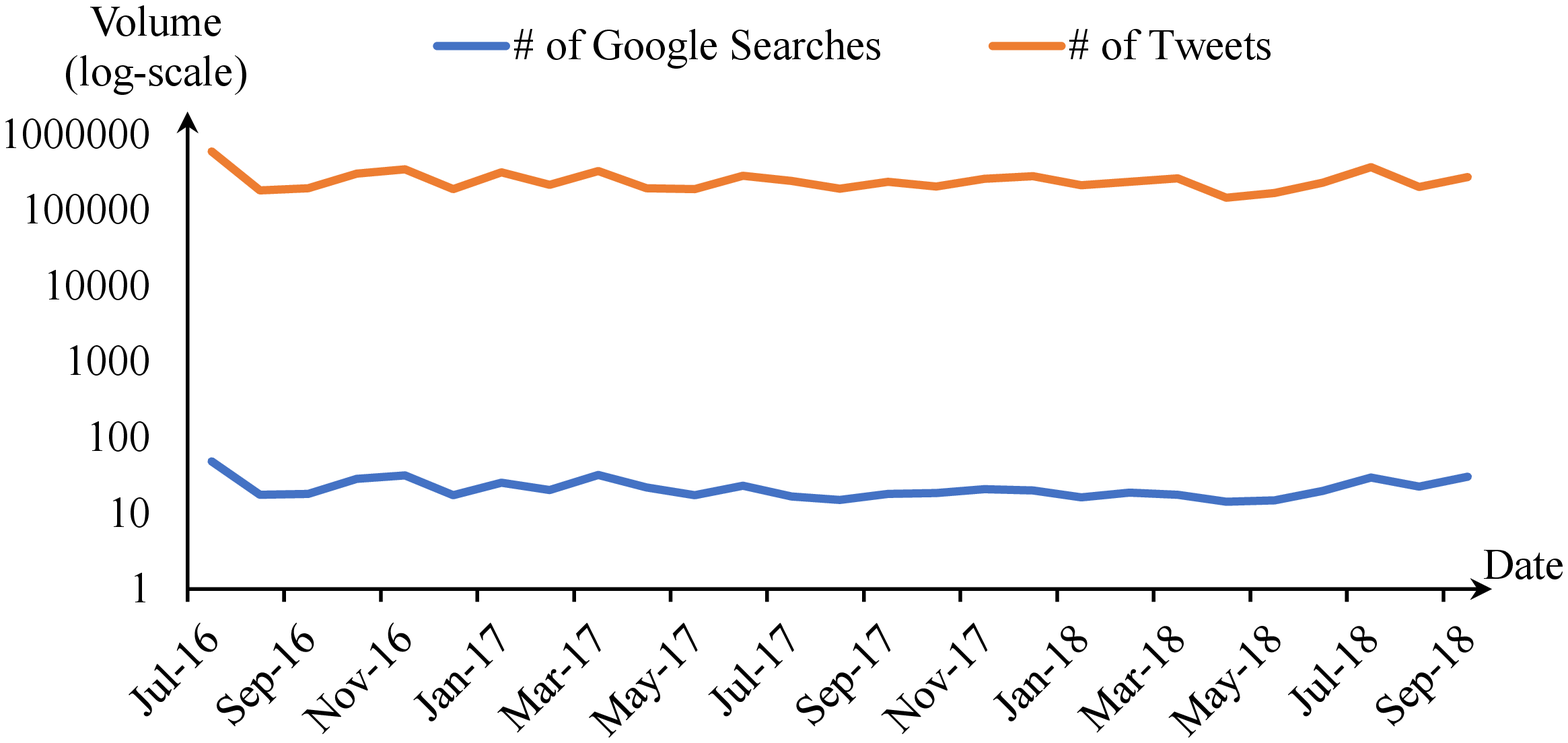}
\captionsetup{justification=centering}
\caption{After the referendum, public interest to Brexit debate continues with the same trend in terms of Twitter posts and Google searches (Correlation 0.92)}
\label{fig1}
\end{figure}

The political issue investigated in this study concerns one of the most important political events of recent times, which defines the process of United Kingdom's exit from the European Union (EU), informally named Brexit. On 23 June 2016, the United Kingdom voted to leave the EU, by 51.9\% for Leave, and 48.1\% for Remain side. However, the local and global impacts of the referendum have made the issue a highly active and long-standing discussion well beyond the end of referendum, as seen in the continuity of the Google search trend (Fig.\ref{fig1}). Another indicator of the constant interest in the subject is the continuing discussion on Twitter, whose trend is highly correlated to the Google search trend on the topic (Pearson correlation of 0.92). This result makes Twitter a convenient data source to analyze the Brexit phenomenon with respect to various aspects.

In this study, we aim to address the following research questions: 

\textbf{Question 1:} Can we determine the political stance of Twitter users with respect to Brexit based on the content they share? Can we analyze how stance evolves in time? 

\textbf{Question 2}: What are the main discussion topics, what is the general sensitivity on these issues and which polarized side reacts to the different issues?

\textbf{Question 3}: Which politicians have been discussed most, what is the general sensitivity with respect to these politicians and which polarized side is more responsive to them? 

\textbf{Question 4}: What is the impact of automated bot accounts to the online discussions, and to which side are they aligned most?

The rest of the paper is organized as follows: We first present the primary studies in our research  focus. Then we give detailed information about our collected data and findings including user demographics and public interest to tweets. We present our two-fold stance classification approach and experiment on the Brexit referendum. Next, we analyze the Twitter accounts regarding their bot behavior, and then we interpret the stance of bot accounts for the Brexit experiment. In topic analysis part, we share the results of topic discovery implementation concerning the public attitude and sentiment of discovered topics. Additionally, we share our findings of engagement of social media users with the politicians, and we analyze the public reaction to these accounts over time. We conclude our work by providing the technical details of our implementation, and with the detailed tables in the appendix section.

\section{Related Work}
\subsection{Social Media and Politics}

Social media has an essential role in terms of sharing information during political happenings. In a study related to German federal elections, the authors found that Twitter is used intensively for political deliberation \cite{Tumasjan-2011}. In another study related to 2013 Italian elections, Vaccari et al. demonstrated that the political deliberation on social media also makes people more conscious and active on the political news \cite{Vaccari-2015}. In a recent study on the Brexit referendum, it is argued that social media data could be used to elucidate the underlying themes/concerns of the political discourse \cite{Khatua-2016}. The intense use of social media in politics makes this platform a vast source of information for understanding various aspects of human behavior and political facts.

\subsection{Stance Classification}

In recent years, the researchers have shown great interest to estimate public opinions about political phenomenon through social media data. Even though there exist some studies \cite{Metaxas-2011} arguing that social media could not be used as a source for electoral predictions in general, several studies achieved notable results. Identifying the users who are in favor of, against, or neutral towards a target is known as stance classification. The target of the stance analysis may be a person, an organization, a government policy, a movement, a product, and so on. On the other hand, stance classification is usually confused with sentiment detection. According to \cite{Mohammad-2017}, while in sentiment analysis the goal is to extract the sentiment from a piece of text, in stance classification the purpose is to determine favorability toward a given (pre-chosen) target of interest. The examples in Table \ref{table1} show the difference; tweets may have the same stance, but opposite sentiment.

Tweets have by nature a concise text structure, which makes the stance classification task more challenging. To overcome this obstacle, many studies have focused on the different steps of machine learning pipeline. For the data annotation part of the supervised learning task, manual \cite{Addawood-2017, Grcar-2017, Celli-2016} or automatical \cite{Lopez-2017} methods have been used. Besides, there also exist some specific studies presenting richer datasets in order to define a gold standard \cite{Hurlimann-2016}. Specifically for Twitter, various feature engineering techniques are implemented such as lexical (n-grams), word-embedding \cite{Mohammad-2017}, syntactic (sentiment, grammatical) \cite{Somasundaran-2009, Tumasjan-2011},  meta-data (retweet count, follower count, mentions), network-specific (retweet-based propagation)\cite{Rajadesingan-2014} and argumentative analysis(argumentativeness, source type) \cite{Addawood-2017}. As a machine learning algorithm, the authors achieved successful results with Naive Bayes\cite{Addawood-2017, Rajadesingan-2014}, Support Vector Machines (SVM) \cite{Mohammad-2017, Addawood-2017, Grcar-2017}, Decision Trees \cite{Addawood-2017} and Recurrent Neural Network (RNN) \cite{Zarrella-2016}, and a combination of RNN with long-short memory (LSTM) and target-specific attention extractor \cite{Du-2017}.

As a complementary step of stance classification, some studies also have applied an age-adjustment since Twitter users do not represent the demographics of voters genuinely. In a recent study, Grcar et al. argue that the age correction changed their prediction outcome from Remain to Leave, by achieving a very close ratio compared to referendum outcome \cite{Grcar-2017}. In another study, Lopez et al. achieved 71\% correlation for Leave and 65\% for Remain without applying any age adjustment \cite{Lopez-2017}. 

\begin{center}
\begin{table*}[t]
\small
\captionsetup{justification=centering}
\caption{Stance and sentiment analysis are separate tasks, expressions from a specific stance may have opposite sentiment.} 
\centering
\begin{tabular}{lll}
\toprule
Tweet&Stance&Sentiment\\
\midrule
I voted \#Remain in the \#Referendum I love my \#European brothers and sisters& Pro\hyp{}Remain & Positive \\
\#Brexit consequences just seem to get worse and worse& Pro\hyp{}Remain & Negative\\
Congratulations, Great Britain on \#Brexit Independence day Enjoy& Pro\hyp{}Leave & Positive \\
I voted leave because I dont think the \#EU works i dont see anything to suggest that it ever will \#Brexit& Pro\hyp{}Leave & Negative \\

\bottomrule
\end{tabular}
\label{table1}
\end{table*}
\end{center}
\subsection{Role of Automated Accounts (Bots) in Elections and Referendums}
While social media is a platform made for the use of people, it is also known that a large share of accounts are automated generators of posts and other activities on social networks. These accounts are often referred to as bots. A type of bots is political social media bots specializing in political issues that are particularly active in public policies, elections, and polarized political discussions \cite{HowardK-2016}. However, their presence in online political discussions could be harmful in many senses. Ratkiewicz et al.'s work on 2010 US Midterm elections and Metaxas et al.'s work on 2010 Massachusetts special election showed that political bots might artificially inflate support for a political candidate \cite{Metaxas-2011, Ratkiewicz-2011}. In a recent study on 2016 US Elections, Bessi and Ferrara found that bot accounts generated about one-fifth of the entire conversation, and their presence negatively affected democratic political discussion rather than improving it, which in turn could potentially alter public opinion and endanger the integrity of the Presidential elections \cite{Bessi-2016}. Similarly, in the case of Brexit referendum, political bots profoundly dominated Twitter for spreading information supporting the idea of leaving the EU, and they generated almost one-third of all content \cite{HowardK-2016}. Again in Brexit referendum, Bastos and Mercea uncovered a bot network comprising 13,493 accounts that massively retweeted user-generated hyperpartisan news and then disappeared from Twitter shortly after the day of the referendum \cite{Bastos-2017}. These studies prove that political bots play an active role in political phenomena and their presence may have negative impacts on the voting results and public opinion.

\subsection{Topic Discovery}
With the high amount of people participating in online social discussions, it becomes challenging to track the discussed topics. For this reason, applying the automatic methods of topic discovery could be an efficient way to explore the discussion focus. Chinnov et al. summarizes the challenges of dealing with short social media texts in topic discovery practices \cite{Chinnov-2015}. As a specific example to solve these problems, Hong and Davison follow an aggregation strategy to increase the amount of short text content for training the topic models \cite{Hong-2010}. As an example to topic discovery applications in particular political science domain, the authors employed US presidential elections and Brexit referendum by creating a general framework based on latent topic models and user features \cite{Guimaraes-2017}. As a baseline of their topic discovery method, they used the algorithm suggested by Zhao et al. \cite{Zhao-2011} which is an adaptation of the Latent Dirichlet Allocation model. In another study \cite{Song-2014}, the authors examined how social and political topics are related to the South Korean presidential elections of 2012, and they had a two-fold method: First, to implement a temporal LDA to analyze and validate the relationship between topics, and then to develop the term co-occurrence retrieval technique in order to compensate LDA's limitations.

\section{Data Collection and Analysis}
\begin{figure}
\centering
\includegraphics[clip, trim=0cm 0cm 0cm 0cm, width=2.2in]{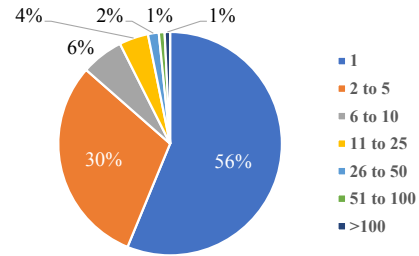}
\captionsetup{justification=centering}
\caption{Tweet post frequency of Users related to Brexit}\label{fig2}
\end{figure}

In our study, we queried for the tweets containing the keyword \textit{Brexit} posted between January 2016 and October 2018. Although the meaning of \textit{Brexit} is UK's exit from the EU, the neutrality of this term has been proven by empirical studies \cite{Lopez-2017}. By using Twitter's API, we collected 10 million tweets sent by 1.5 million users in different languages. As shown in Fig.\ref{fig2}, more than half of the users participated in the discussion only once.
\begin{figure*}[t]
\centering
    \begin{subfigure}[t]{1.5in}
        \includegraphics[clip, trim=0cm 2cm 0cm 0cm, width=1.5in]{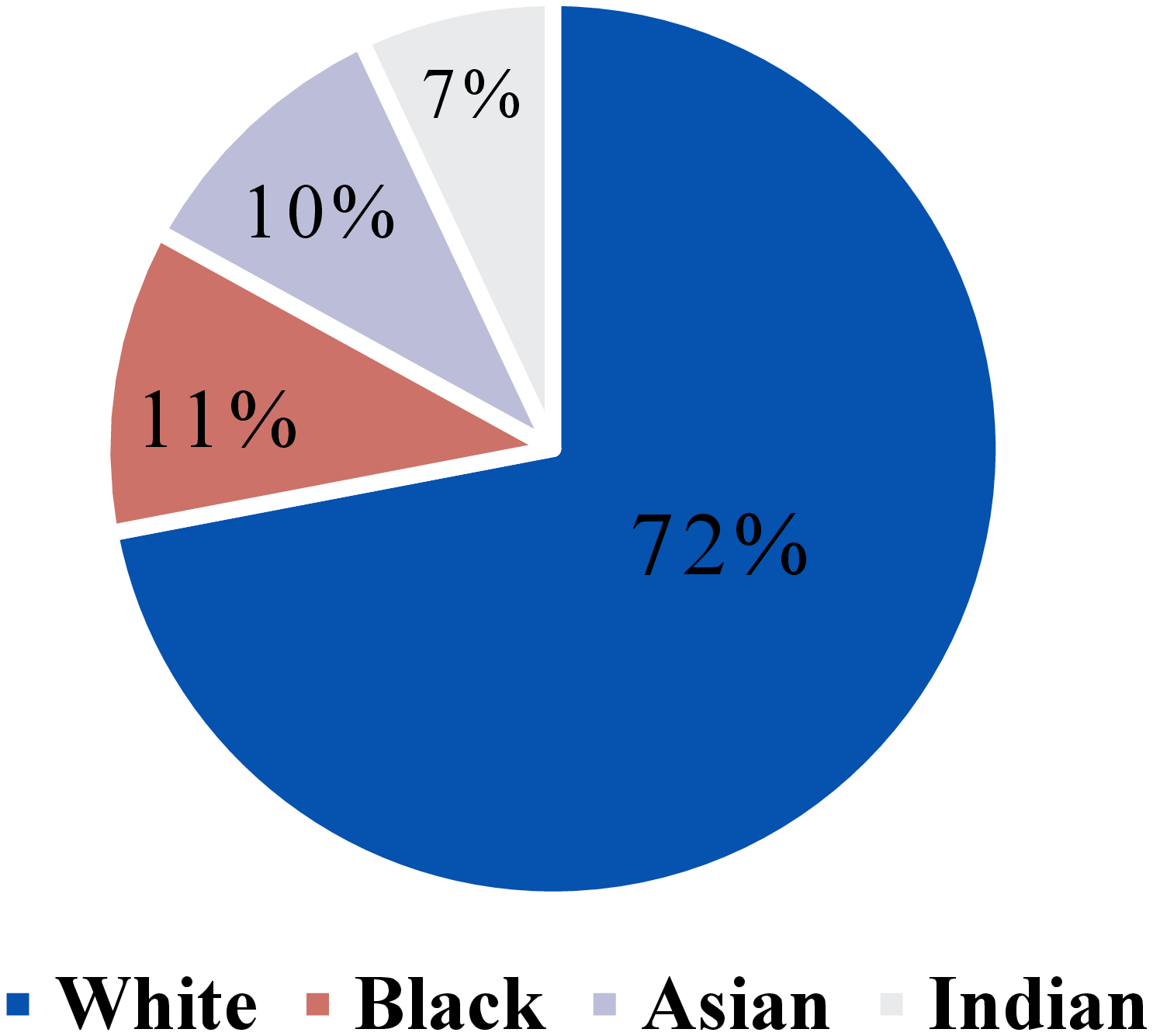}
        \captionsetup{justification=centering}
        \caption{Users by Ethnicity}\label{fig:fig3a}
    \end{subfigure}
    \begin{subfigure}[t]{1.5in}
        \includegraphics[clip, trim=0cm 2cm 0cm 0cm, width=1.5in]{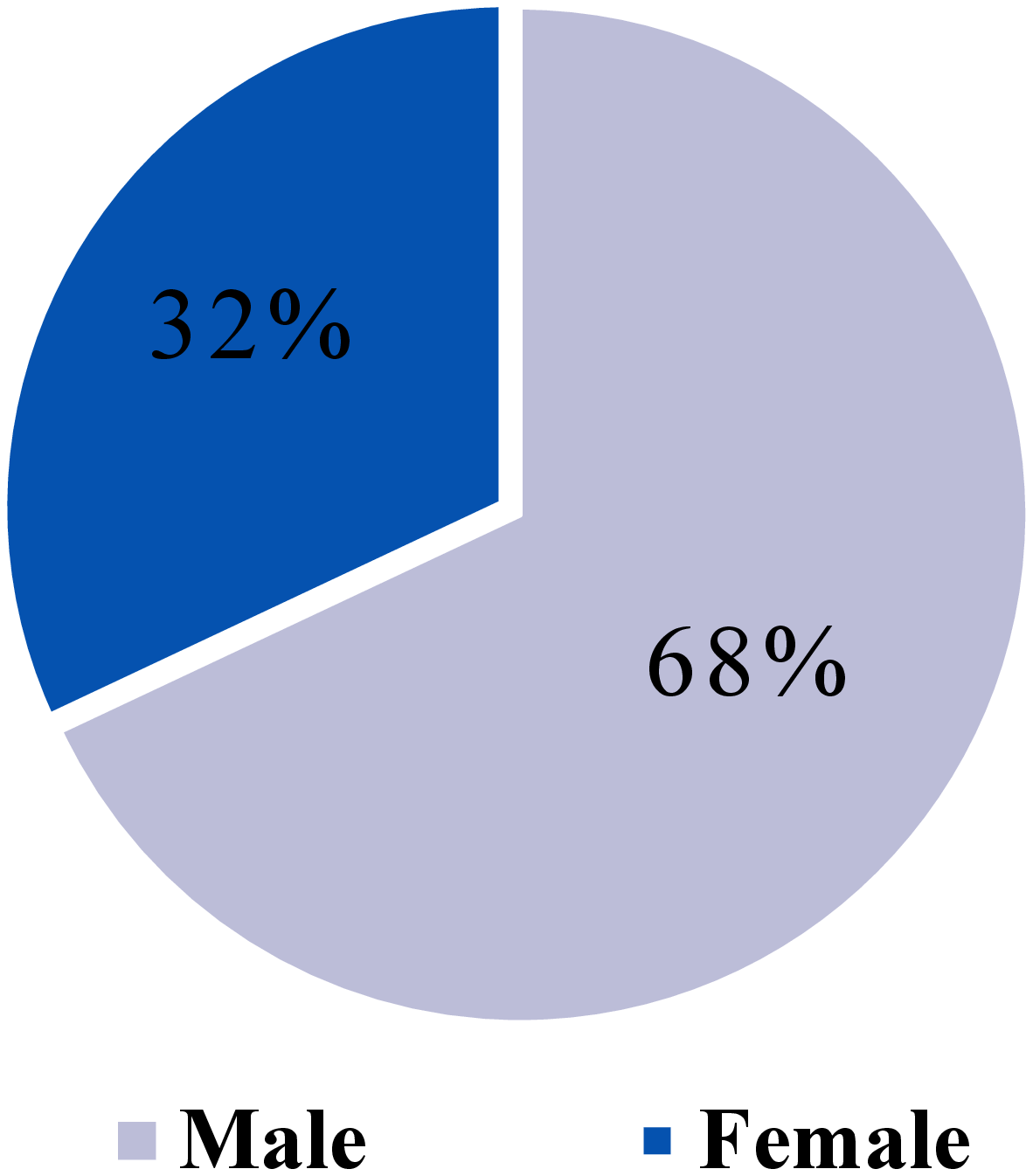}
        \captionsetup{justification=centering}
        \caption{Users by Gender}\label{fig:fig3b}
    \end{subfigure}
    \begin{subfigure}[t]{1.5in}
        \includegraphics[clip,width=1.5in]{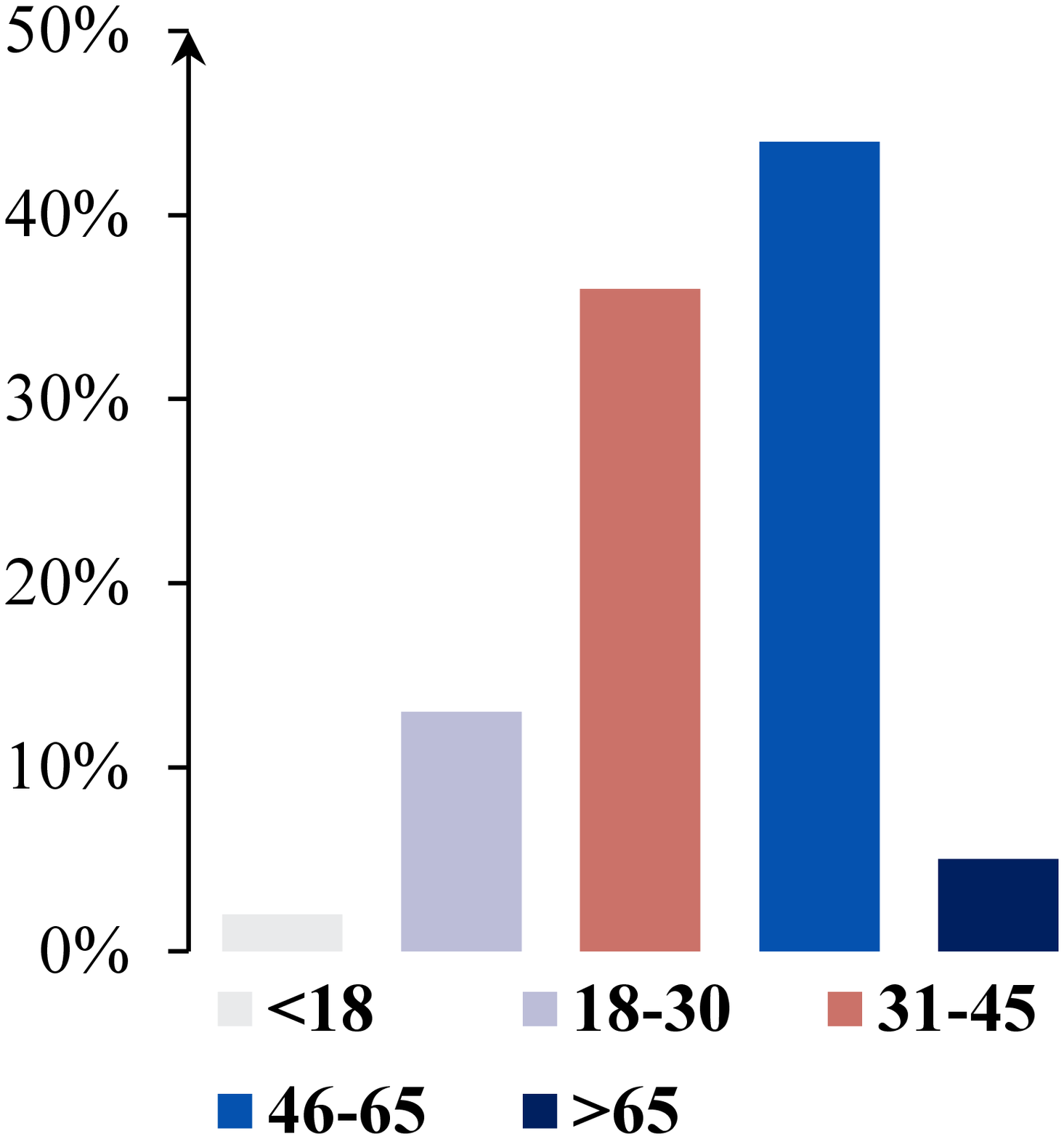}
        \captionsetup{justification=centering}
        \caption{Users by Age}\label{fig:fig3c}
        \end{subfigure}
    \begin{subfigure}[t]{1.5in}
        \includegraphics[clip, trim=0cm 2cm 0cm 0cm, width=1.7in]{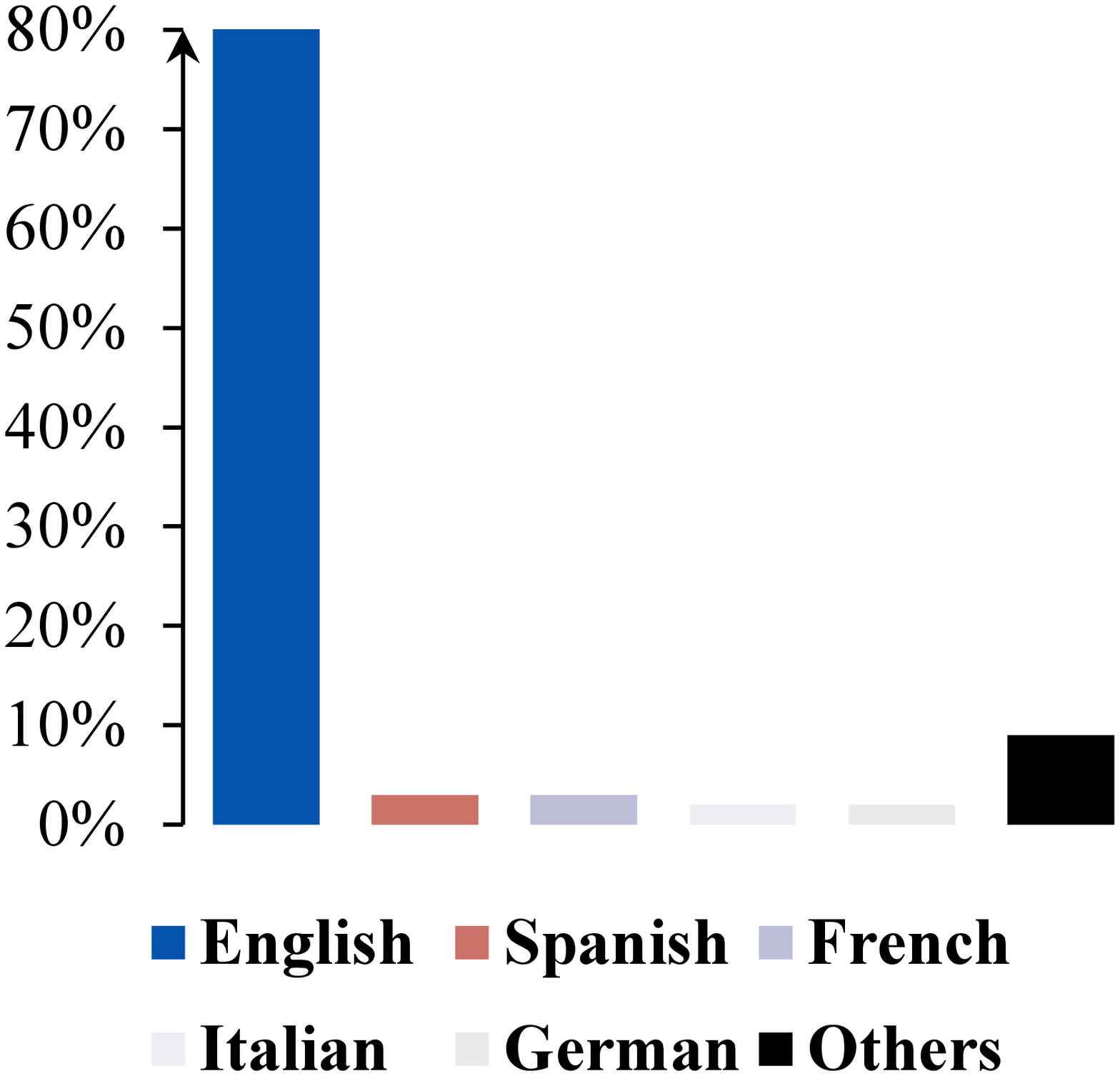}
        \captionsetup{justification=centering}
        \caption{Tweets by Language}\label{fig:fig3d}
    \end{subfigure}
    
    \begin{subfigure}[t]{3in}
        \includegraphics[clip,trim=0cm 0cm 0cm 0cm, width=3in]{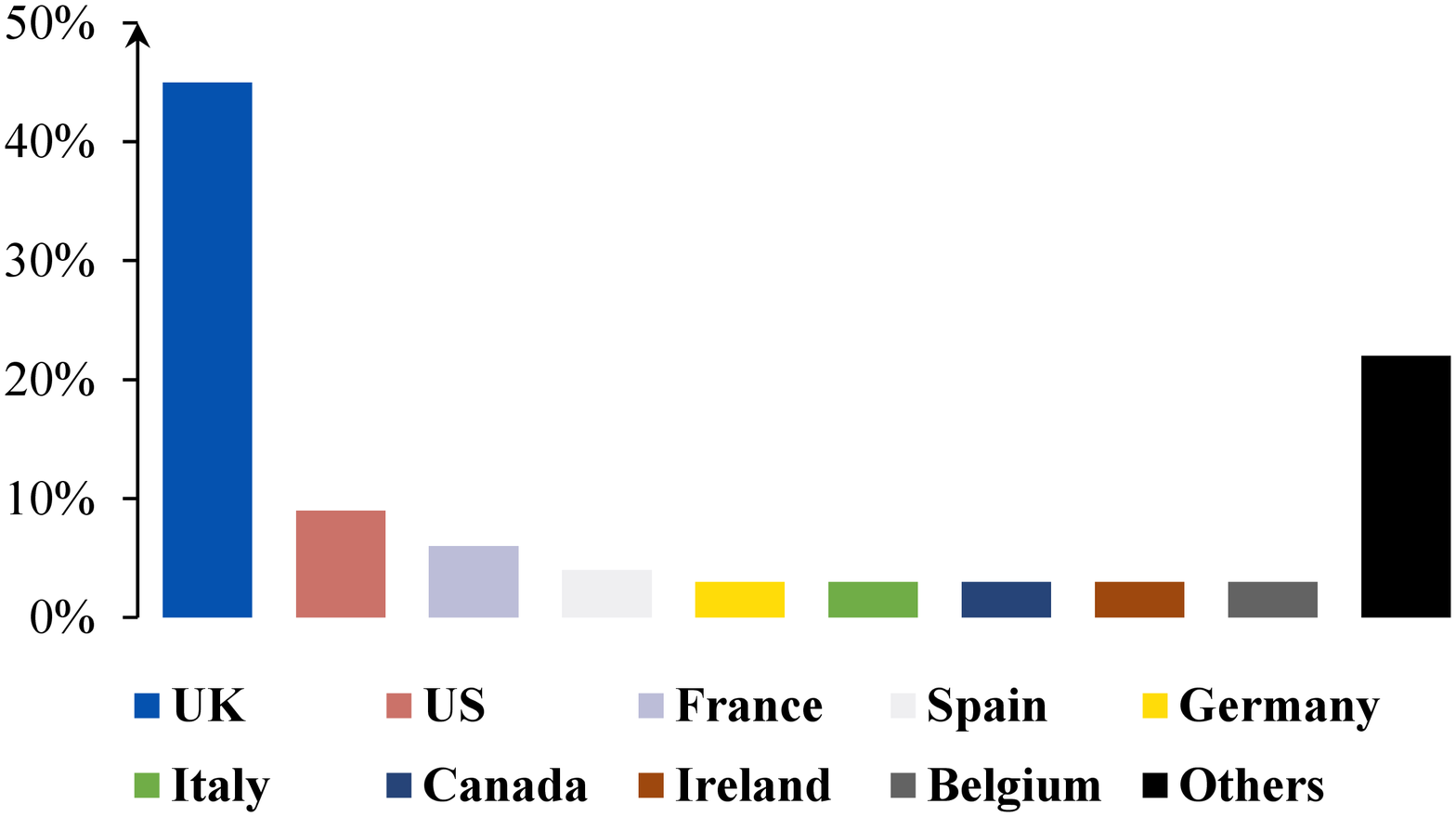}
        \captionsetup{justification=centering}
        \caption{Tweets by Country}\label{fig:fig3e}
    \end{subfigure}
    \begin{subfigure}[t]{3.2in}
        \includegraphics[clip, trim=0cm 0cm 0cm 0cm, width=3in]{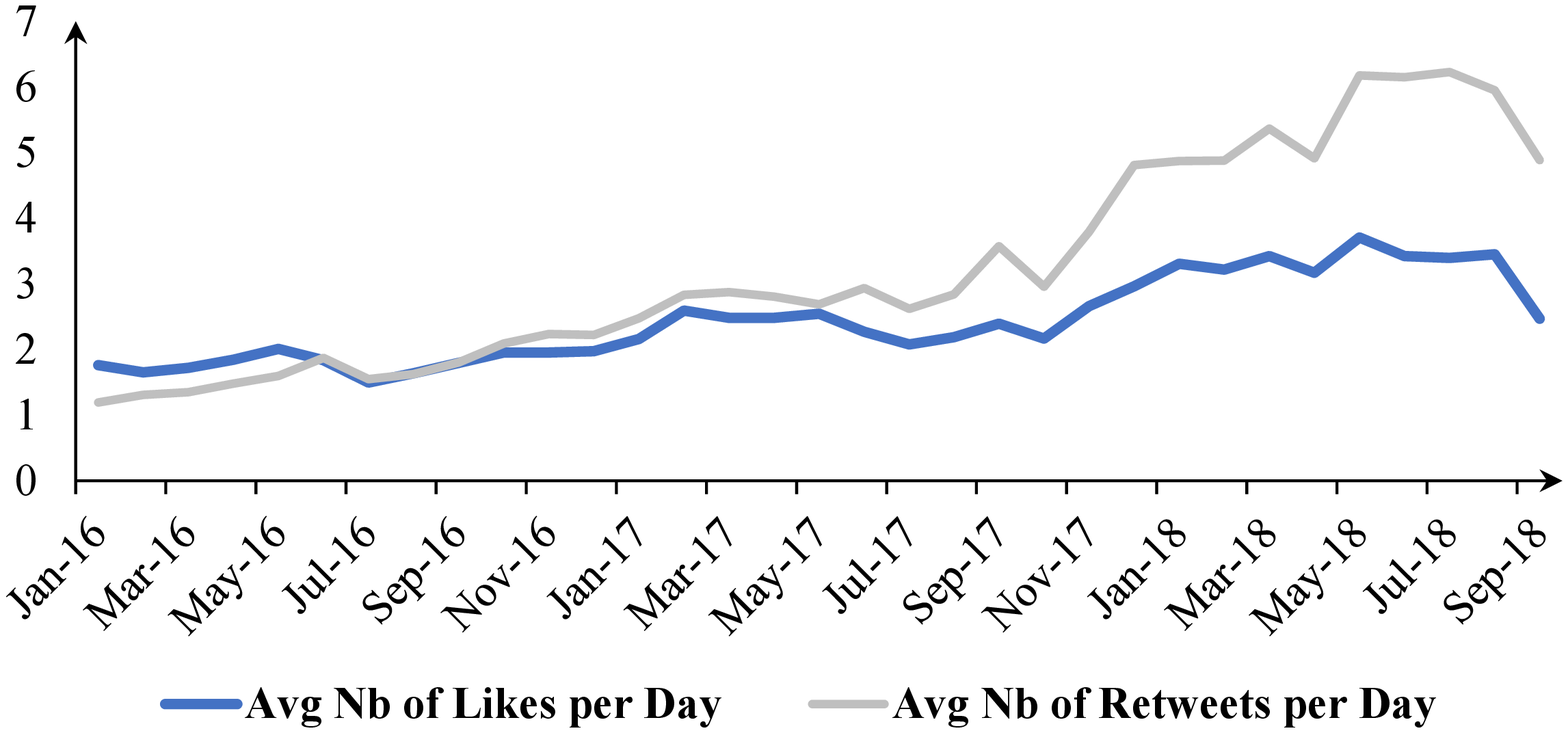}
        \captionsetup{justification=centering}
        \caption{Impact of Tweets based on Nb. of Retweets and Likes}\label{fig:fig3f}
    \end{subfigure}
\captionsetup{justification=centering}
\caption{User demographics and tweet meta-data analysis}\label{fig3}
\end{figure*}
\subsection{User Demographics, Spatial Analysis}
Social media messages may contain additional attributes that may provide demographics and location information about users. In our approach to demographic analysis, we benefited from the profile photos of Twitter users. Taking into account Jung et al.'s experiments, we analyzed profile images through face detection and recognition in order to find the age, gender and ethnicity of users with a single face in the profile photo \cite{Jung-2017}. According to our analysis, 30\% of the user base has a single face in profile photos, and we have been able to make demographic inferences for that user base. Our results showed that users of every ethnic background share their opinions on the Brexit process (see Fig.\ref{fig:fig3a}). On the other hand, the percentage of male users is slightly higher than the Twitter average (see Fig.\ref{fig:fig3b}) \footnote{Statista \url{https://www.statista.com/statistics/828092/distribution-of-users-on-twitter-worldwide-gender/}}. 

Surprisingly, we have found that young people are less interested in the Brexit debate. Although 37\% of Twitter users are under 18 years old according to the latest statistics\footnote{Omnicore Agency  \url{https://www.omnicoreagency.com/twitter-statistics/}}, this ratio is only 15\% in our database (see Fig.\ref{fig:fig3c}). This result is important because in some of the Brexit related stance classification studies \cite{Grcar-2017}, the authors performed age adjustments on their prediction results by claiming that the Twitter users are much younger than English voters. However, our result shows that the participants to Brexit debate on Twitter do not represent general Twitter users.

In our language and spatial analysis, we found that 81\% of tweets are written in English (Fig.\ref{fig:fig3d}), and 45\% of tweets are posted from the United Kingdom (Fig.\ref{fig:fig3e}). In the stance classification and topic discovery analyses where the textual content is the main feature, we only use the tweets written in English.

\subsection{Tweet and User Meta-Data Analysis}
In this section, we provide useful insights based on our meta-data analysis on the Twitter users and their posts. The first valuable information we found is that the average number of followers of Twitter users participating in the Brexit discussions is six times higher than the average Twitter user average, which could be interpreted as the audience discussing Brexit is composed of highly influential people.\footnote{DMR Business Statistics \url{https://expandedramblings.com/index.php/march-2013-by-the-numbers-a-few-amazing-twitter-stats/}}
\begin{table}
\scriptsize
\captionsetup{justification=centering}
\caption{Stance-Indicative (SI) and Stance-Ambiguous (SA) Hashtags}
\centering
\begin{tabular}{p{1cm} p{6cm} }
\toprule
Stance&Characterizing Hashtags\\
\midrule
\texttt{Remain}& \#strongerin, \#voteremain, \#intogether, \#labourinforbritain, \#moreincommon, \#greenerin, \#catsagainstbrexit, \#bremain, \#betteroffin, \#leadnotleave, \#remain, \#stay, \#ukineu, \#votein, \#voteyes, \#yes2eu, \#yestoeu, \#sayyes2europe, \#fbpe, \#stopbrexit, \#stopbrexitsavebritain\\
\hline
\texttt{Leave} & \#leaveeuofficial, \#leaveeu, \#leave, \#labourleave, \#votetoleave, \#voteleave\#takebackcontrol, \#ivotedleave, \#beleave, \#betteroffout, \#britainout, \#nottip, \#takecontrol, \#voteno, \#voteout, \#voteleaveeu, \#leavers, \#vote\_leave, \#leavetheeu, \#voteleavetakecontrol, \#votedleave \\
\hline
\texttt{Ambigious} & \#euref, \#eureferendum, \#eu, \#uk\\
\bottomrule
\end{tabular}
\label{table2}
\end{table}
Our second finding shows that Twitter users become more interested in Brexit-related content in time, even more than in the day of the referendum. Figure \ref{fig:fig3f} illustrates the increase in the number of retweets and likes per tweet over time.
\section{Brexit Stance Classification}
In stance classification, we aim to find users in pro\hyp{}Remain or pro\hyp{}Leave stance and analyze their participation in the Brexit discussions. Some studies \cite{Grcar-2017, Lopez-2017} considered the presence of stance-indicative (SI) hashtags as an effective way to discover polarized tweets and users. The disadvantage of using this method is that it cannot evaluate tweets that do not contain SI hashtags. Unfortunately, this typically includes a substantial share of tweets. The solution we propose is to divide our dataset into two subsets, the ones that contain SI hashtags and the ones that don't. Then, we classify the tweets with SI hashtags by rule-based method, and the remaining tweets by machine learning methods. Notice that in our context, only 8\% of the tweets contain SI hashtags. Thanks to our approach, we can instead analyze the remaining 92\% too. After classifying each tweet as pro\hyp{}Remain, pro\hyp{}Leave or non-polarized, we will be able to determine each user's stance by looking at the number of tweets in each class.
\subsection{Rule-based Classification}
Hashtags are commonly used by Twitter users to express their stance in a political phenomenon. According to our analysis, between January 2016 and September 2018, more than 600 thousand unique hashtags were used with the Brexit hashtag. As shown in Table \ref{table2}, we created a list of stance-indicative (SI) and stance-ambiguous hashtags by finding the most commonly used hashtags and considering the findings of other Brexit related studies. In this method, we classified the tweets based on the following hypothesis.
In our approach, the stance of a tweet is:
\vspace{-1.7mm}
\begin{itemize}
\item \textit{Pro\hyp{}Remain (PRT)}, if it contains at least one \textit{Remain}, but not any \textit{Leave} related hashtag,
\item \textit{Pro\hyp{}Leave (PRL)}, if it contains at least one \textit{Leave}, but not any \textit{Remain} hashtag,
\item \textit{Non-polarized} for all other cases.
\end{itemize}
Then, to calculate the user stance, we applied the following formula considering all tweets of the user in our database.
\vspace{-0.2mm}
\small
\begin{align*}
&Score=\frac{\sum PRT}{\sum{PRT} + \sum{PRL}}\\
\vspace{2mm}
&UserStance=\left\{{\begin{array}{ll}
             Pro\hyp{}Leave, \qquad if \quad Score  \quad <\: 0.4 \\
             Pro\hyp{}Remain, \quad if \quad Score\quad >\: 0.6 \\
             Non-polarized,\quad otherwise
             \end{array}}\right.\\
\end{align*}
    
In our comparative approach, we only take into account Pro\hyp{}Leave and pro\hyp{}Remain users, and we get the ratio of a class by dividing its value to the sum of two classes. As a result, we found that the number of pro\hyp{}Remain users is relatively higher than the number of pro\hyp{}Leave users (see Table \ref{table3}). However, this method classified 92\% of tweets as non-polarized because they do not contain SI hashtags. Within our knowledge, Twitter has become the primary place for online social discussions on the Brexit referendum, and there should be a higher number of active polarized users on Twitter. Therefore, we have developed the following complementary method using machine learning techniques for stance classification of the tweets not featuring SI hashtags. 

\subsection{Machine Learning (ML) Based Classification} 
In this task, we only focused on the tweets that are labeled as non-polarized in the previous method. For the preparation of training and development set for our learning-based classifier, a subject expert involved in our study, and prepared three sets of 1000 tweets from each class: \textit{pro\hyp{}Remain}, \textit{pro\hyp{}Leave} and \textit{non-polarized}. In terms of feature engineering, we normalized the tweets with a Twitter-specific tokenizer and then transformed to n-gram pairs (uni-bi-trigrams). For the implementation of the classification algorithm, we tested various algorithms, and we obtained the best results with the Support Vector Machines having a linear kernel. In a recently shared task about stance classification, Mohammd et al. obtained the highest score among other tasks with a machine learning model similar to ours \cite{Mohammad-2017}. 
    
As a result of the 10-fold cross verification, the weighted average F1 score and AUC scores achieved to 0.71 and 0.80. By predicting the tweets using this model, we obtained 2.1 million tweets from pro\hyp{}Remain and 1.8 million tweets from pro\hyp{}Leave classes. Then, for the validation of the classification task, a subject expert evaluated the predicted labels on a randomly selected subset of data. As a result, we found that the model's variance is less than 5\% for both classes.

This method allowed us to detect a significant amount of polarized tweets. In the final step, we obtained a complete tweet set of 2.55 million pro\hyp{}Remain and 1.8 million pro\hyp{}Leave tweets by combining the results of rule-based and machine learning-based methods. Over this dataset, we applied the user stance evaluation, and we found that 432,000 users are pro\hyp{}Remain and 309,000 of users are pro\hyp{}Leave.

\subsection{Analysis of Changes of Users' Stance in time}
\begin{figure}
\centering
\includegraphics[clip, trim=2cm 0cm 0cm 8cm, width=3in]{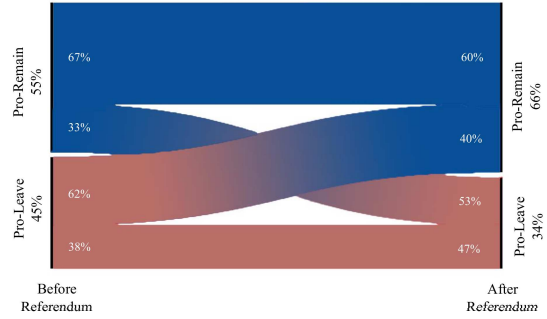}
\captionsetup{justification=centering}
\caption{When we compare the users before and after the referendum, we see that the change of stance is mostly from pro\hyp{}Leave to pro\hyp{}Remain.
} 
\label{fig4}
\end{figure}

Besides the static classification of users' stance, we also analyzed the change in stance from two perspectives. In our first approach, we compared the users' pre and post-referendum tweets, and we found that the number of users who change their stance is significantly higher in the pro\hyp{}Leave side (62\%)  than the pro\hyp{}Remain side (33\%) (Fig.~\ref{fig4}). 

\begin{figure}[t]
\centering
\includegraphics[clip, trim=0cm 0cm 0cm 0cm, width=3in]{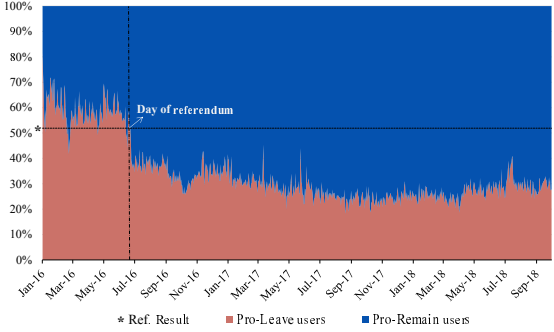}
\captionsetup{justification=centering}
\caption{Our daily analysis identifies a major change in stance after the referendum. In the following time, the rate of participation in the pro\hyp{}Remain side was consistently higher than Pro\hyp{}Leave.} 
\label{fig5}
\end{figure}

In our second approach, we analyzed monthly changes in the stance of users. By calculating a single stance value for users from their monthly tweets, we visualized the increases and decreases of participation to debate from each side (Fig.~\ref{fig5}). Our result validates the referendum outcome with 51\% of pro\hyp{}Leave and 49\% pro\hyp{}Remain users. Furthermore, our results show that the percentage of Pro\hyp{}Remain users is varying between 60\% and 70\% over the past two years.

\begin{table}[t]
    \small
    \centering
    \captionsetup{justification=centering}
    \caption{Two-fold approach to classify the user stance}
    \begin{tabular}{llll}
    \toprule
    Method & Type & Remain & Leave\\
    \midrule
    Rule-based (RB) & Tweets  & 462K & 254K \\  
    {} & Users  & 62K & 38K \\  
    \midrule
    Machine Learning & Tweets &  2.1M & 1.8M \\
    based (MLB) & Users  & 408K & 296K \\  
    \midrule
    Merged (RB + MLB) & Tweets & 2.56M & 2.05M\\
    {} & Users & 432K & 309K \\
    \bottomrule
    \end{tabular}
    \label{table3}
\end{table}

\section{Impact of Bots on Online Social Debate and Overall Stance}
As we described in the Related Work section, various studies show the relevance of political bot accounts during political elections and referendums. In a recent article \cite{Howard-2018}, the author states that the computational propaganda powered by political bots takes many forms: networks of highly automated Twitter accounts; fake users on Facebook, YouTube, and Instagram; chatbots on Tinder, Snapchat, and Reddit. These bot accounts track different strategies to mimic human users, making it difficult for social media providers to identify them. In our Brexit experiment, we found that there are many accounts deactivated or suspended accounts. On the other hand, we found that many Twitter bot accounts are still alive. As a method of identifying Twitter bot accounts, we benefited from the state\hyp{}of\hyp{}the\hyp{}art bot detector which assigns a bot score to a Twitter account in the range (0,1) describing how likely it is to be an automated account with 1 being the maximum probability \cite{Davis-2016} \footnote{Botometer \url{https://botometer.iuni.iu.edu/}}. As suggested by the author, we mark an account as bot if it's score is higher than 0.8. As a result of our analysis, we found that the percentage of bot accounts that are still alive on Twitter is 2.2\%, and their average post frequency was 25\% higher than the non\hyp{}bot accounts. Our result confirms the statement of Howard and Kollanyi \cite{HowardK-2016}, claiming that the bot accounts were highly active in the Brexit debate.

\begin{figure}[t]
\centering
\captionsetup{justification=centering}
\includegraphics[clip, trim=0cm 0cm 0cm 0cm, width=3in]{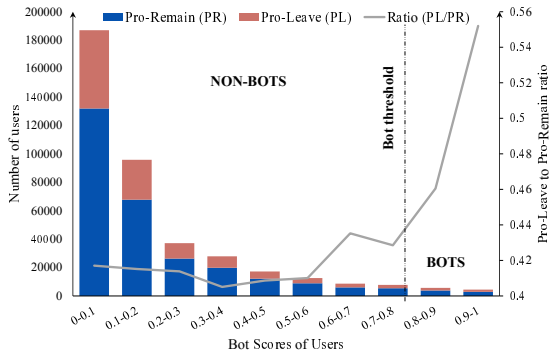}
\caption{The higher the bot score for a Twitter account, the more likely of being in pro\hyp{}Leave stance} 
\label{fig6}
\end{figure}

By extending our findings one step further, we combined the bot scores with the results of user stance classification described in the previous section. Interestingly, our result shows that the higher the bot score, the more likely the account is in a pro\hyp{}Leave position (See Fig.\ref{fig6}).

\section{Topic Discovery}
We analyzed the topics of Brexit\hyp{}related discussions on Twitter. Brexit is a long\hyp{}term happening regarding its impact on society; therefore a variety of topics have been discussed by Twitter users in the context of Brexit including immigration, borders, and economic impacts. In our study, we benefited from Latent Dirichlet Allocation (LDA) algorithm to extract the topics \cite{Blei-03}. One questionable aspect of applying LDA algorithm for our scenario could be the shortness of text contents and data ambiguity. To overcome this limitation, we applied a data selection strategy to eliminate the shortest and non\hyp{}influential tweets. As a result, we executed the topic discovery algorithm on a dataset containing 306 thousand tweets posted between January 2016 and October 2018. We evaluated the LDA algorithm based on coherence score and subject expert feedback. We didn't use the perplexity score because perplexity and human judgment are often not correlated, and even sometimes slightly anti\hyp{}correlated \cite{Chang-2009}. 

\subsection{Full\hyp{}period Topic Analysis}
In our first experiment, we directly fed the LDA algorithm with the whole set of 306 thousand tweets on January 2016 and October 2018. Then, our subject expert assigned labels to the discovered topics through the representative words as shown in Table \ref{table4}. However, we found that the quality of the topics is not high in terms of both the coherence score and the subject expert evaluation. This is mainly because the model is ineffective in finding time\hyp{}varying topics as it operates over a long period. This has led to the inability to find short\hyp{}term but essential issues. 

\begin{figure}[t]
\centering
\includegraphics[clip, trim=0cm 0cm 0cm 0cm, width=3.4in]{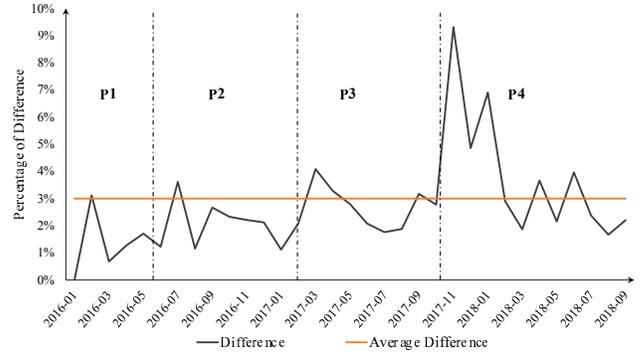}
\captionsetup{justification=centering}
\caption{Monthly cumulative change of topics based on full\hyp{}period analysis}\label{fig7}
\end{figure}

For this reason, we decided to reduce the time interval, and we did this systematically by examining the change in the participation to the topics on a monthly basis. As shown in Figure \ref{fig7}, we found that the percentages of change were higher in four specific times. Therefore, we applied the LDA algorithm separately with the tweets sent over these periods. In this way, we achieved more consistent and specific topics than our first experiment. 

For instance, our experiment on time period P4 has successfully discovered many key topics related to the cabinet, trade deals with EU, new referendum expectations, Scottish referendum and Irish border (see Table \ref{table5}). Other results are shown in Appendix (Table \ref{table8},\ref{table9}, and \ref{table10}). 
\begin{table}[t]
\captionsetup{justification=centering}
\caption{The experiment on the whole time period contains topics that are not consistent and repetitive, and cannot find discover many key topics in the Brexit process.}
\scriptsize
\begin{center}
\begin{tabular}{ ll }
 \toprule
 Label & Top Representative Words\\
 \midrule 
1.US\hyp{}Russia & trump,farage,attack,threat,russia,boris,putin,usa\\
2.Europe & europe,trust,germany,maymustgo,france,cameron,merkel\\
3.Results & live,ask,watch,question,immigration,war,secretary,maga\\
4.Labour & tory,labour,party,scotland,corbyn,leader,political\\
5.Trade & deal,good,trade,bad,ita,free,dona,agreement,sell,offer\\
6.Inconsist. & explain,medium,consequence,message,send,suggest,letter\\
7.Vote & vote,leave,people,want,remain,british,government,\\
8.Plan & theresa\_may,conservative,plan,borisjohnson,deliver,fail\\
9.Inconsist. & britain,great,world,article,wrong,nation,damage\\
10.Economy & year,job,business,economy,warn,economic,impact,face \\
11.Inconsist. & take,govt,control,law,place,interest,protect,strong,bill\\
12.Remain & stay,customs\_union,single\_market,england,membership\\
13.Economy & nhs,pay,money,tax,fund,save,foreign,spend\\
14.Remain & stop,join,help,stand,thank,pro,fight,speak,march,lord\\
15.Remain & stopbrexit,fbpe,country,right,work,thing,remainer,finalsay\\
16.Leave & lie,campaign,ukip,euref,voteleave,leaveeu,blame,truth \\   
17.Borders & hard,ireland,border,idea,problem,possible,irish,mess\\ 
18.Polarized & nigel\_farage,feel,freedom,disaster,act,remainernow\\
19.Inconsist. & today,new,day,pm,post,talk,eu,look,future,read \\
20.Inconsist. & prime\_minister,reality,everything,westminster,charge\\
\bottomrule
\end{tabular}
\end{center}
\label{table4}
\end{table}

\subsection{Relations Among Topics, User Stance and Sentiment}

By taking the results of the previous section one step further, we also revealed which polarized sides do the sharing of the topics found.(pro\hyp{}Remain / pro\hyp{}Leave), and what is the general sentiment to these topics. Our aim is to generate statements such as: \textit{For the topic related to immigration, mostly the pro\hyp{}Remain/pro\hyp{}Leave users are tweeting, and the overall sentiment to this topic is positive/negative}. In this task, we used our stance classification results and syntactic word\hyp{}based sentiment detection approach.

In our findings, we included the comparison of pro\hyp{}Remain/pro\hyp{}Leave stances and Positive/Negative sentiments for each discovered topic (see Table \ref{table5}). One of the impressive results is that the 97\% of tweets of the \textit{New Referendum Request} topic is from the pro\hyp{}Remain side with a negative sentiment. On the other hand, for the topic entitled \textit{cabinet}, 73\% of tweets are posted by pro\hyp{}Leave side. 88\% of the tweets sent related to the Irish border issue have a positive feeling.
\begin{table}
\begin{center}
\captionsetup{justification=centering}
\caption{Stance and Sentiments of Topics discovered in P4 time period (Nov,2017\hyp{}Sep,2018)} 
\includegraphics[clip, trim=0cm 0cm 0cm 0cm,scale=0.8]{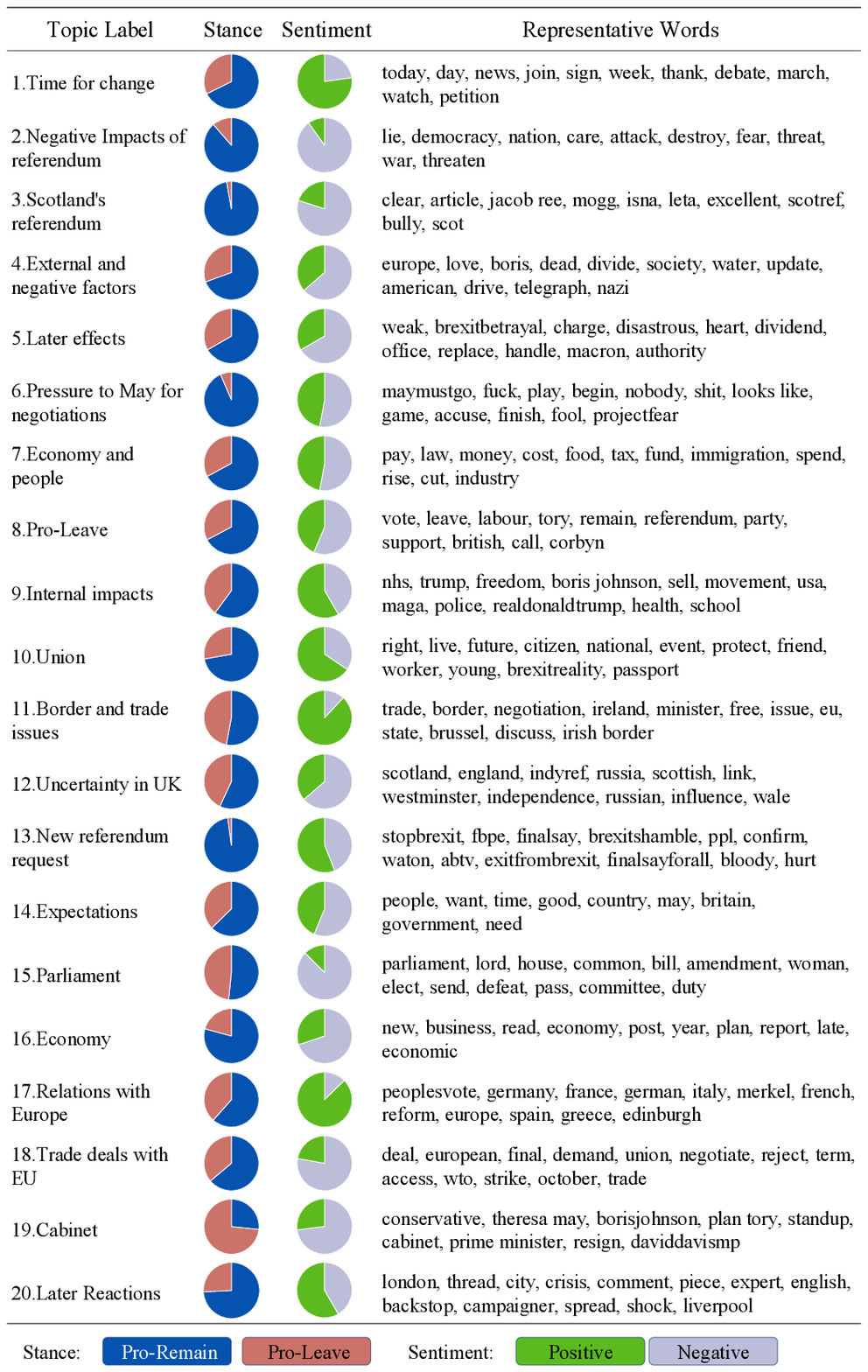}
\label{table5}
\end{center}
\end{table}

\begin{center}
\begin{table}
\scriptsize
\captionsetup{justification=centering}
\caption{Twitter accounts that are mentioned by other users for 10K+ times.The starred accounts have very high bot scores.} 
\centering
\begin{tabular}{lll}
\toprule
Politicians&News Channels&Campaign\hyp{}Party\\
\midrule
\texttt{@theresa\_may}& \texttt{@BBCNews} & \texttt{@UKLabour}\\
\texttt{@jeremycorbyn} & \texttt{@SkyNews}& \texttt{@Conservatives}\\
\texttt{@Nigel\_Farage} & \texttt{@guardian}& \texttt{@LeaveEUOfficial}\\
\texttt{@BorisJohnson} & \texttt{@LBC}& \texttt{\textbf{@vote\_leave*}}\\
\texttt{@realDonaldTrump} & \texttt{@FT}& \texttt{@LibDems}\\
\texttt{@David\_Cameron} & \texttt{@Independent}& \texttt{@UKIP}\\
\texttt{@DavidDavisMP} & \texttt{@Telegraph}& \texttt{\textbf{@StrongerIn*}}\\  
\texttt{@Jacob\_Rees\_Mogg} & \texttt{@afneil}& \texttt{\textbf{@theSNP*}}\\  
\texttt{@Anna\_Soubry} & \texttt{@BBCr4today}& \\
\texttt{@ChukaUmunna} &  \texttt{@MailOnline}&\\
\texttt{@Keir\_Starmer} &  \texttt{@business}&\\
\texttt{@NicolaSturgeon} & & \\
\texttt{@MichelBarnier} & & \\
\texttt{@Andrew\_Adonis} & & \\
\bottomrule
\end{tabular}
\label{table6}
\end{table}
\end{center}
\section{Analysis of Politician Accounts on Twitter}
\begin{figure}[t]
\centering
\includegraphics[clip, trim=0cm 0cm 0cm 0cm, width=3in]{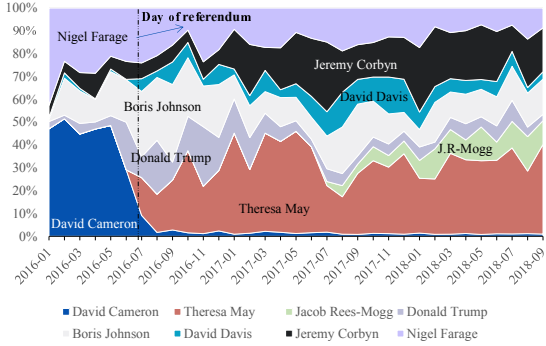}
\captionsetup{justification=centering}
\caption{Comparative influence levels of politicians over time
} 
\label{fig8}
\end{figure}

Online social media is a significant platform for politicians to interact directly with the public. Twitter users can reach politicians directly by mentioning their accounts and declare their opinions. In our study, we analyzed to find the politician accounts that interacted most, and as a result of our categorization through the most frequently mentioned accounts (Table \ref{table6}), we focused on ten politician accounts.
In our comparative temporal analysis (see Fig.\ref{fig8}), we have obtained the following insights:

\begin{itemize}
\item 
James Cameron had lost his influence in Twitter after handing over his Prime Minister (PM) role to Theresa May. New PM Theresa May has become the essential actor of the Brexit process, although she was not known widely by the public before the referendum.
\item 
At the beginning of July 2017, we discovered a sudden increase in Jacob Rees\hyp{}Mogg's influence on Twitter. He increased his popularity and surpassed the Twitter account, Nigel Farage.
\item
After becoming President of the United States, Donald Trump became very popular at the center of the Brexit debate, and this interest continued until 2017. However, as of February 2017, another politician, Jeremy Corbyn, was discussed more than Trump and other politicians.
\end{itemize}

\begin{table}
\begin{center}
\caption{Stance and sentiments of tweets related to specific politician accounts} 
\includegraphics[clip, trim=0cm 0cm 0cm 0cm, scale=0.5]{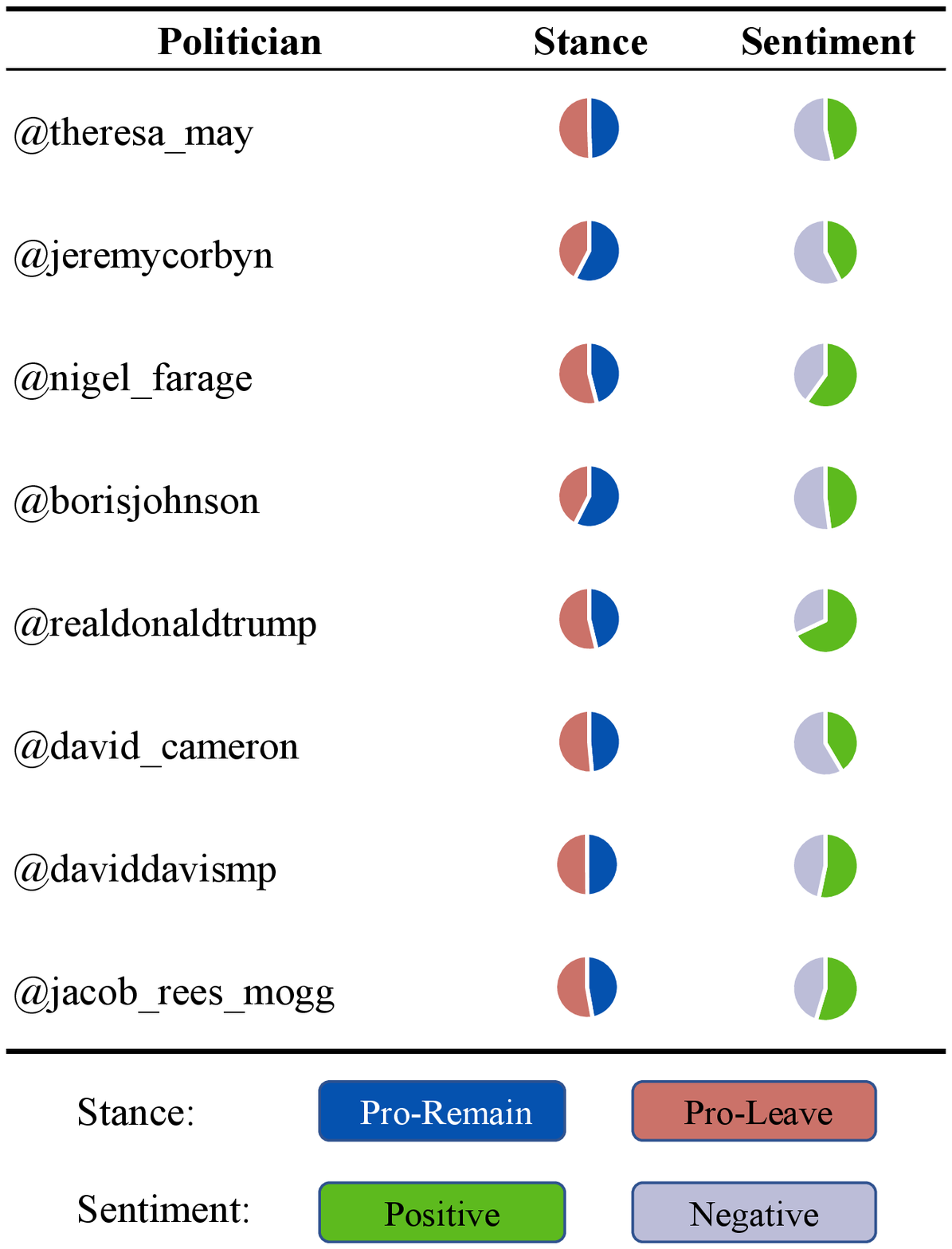}
\captionsetup{justification=centering}
\label{table7}
\end{center}
\end{table}

In addition to the temporal analysis, we also measured the sentiment and stance of Brexit related tweets that are mentioning politician accounts (Table \ref{table7}). The characteristics of mentions to Nigel Farage and Donald Trump is very similar; those tweets are mostly positive and sent by pro\hyp{}Leave users. On the other hand, Jeremy Corbyn and Boris Johnson are mostly discussed by pro\hyp{}Remain users.

\section{Conclusions}
In this study, we provide a comprehensive analysis of the interpretation of large-scale, long-running political phenomena in online social media. By focusing on the one of the most important political happening of recent times, the Brexit referendum, we applied several computational social science techniques on 33 months of public Twitter data. We first performed a demographic analysis on the users participating in the online social discussions on Twitter, and then we predicted their polarized stance with a combination of rule-based and machine learning-based classification methods. As a result of our temporal analysis, we found that the highest change in user stance after the referendum occurred on the pro\hyp{}Leave side. Additionally, we extracted the most significant topics of debate, and we measured the public stance and sentiment in respect to these topics. Finally, we analyzed reactions to public accounts of politicians in stance and sentiment, and we compared the volumetric distribution of reactions over time. As a result of our study, we show that social media\hyp{}based analysis could provide useful insights to understand people and facts during political phenomena.

\section{Implementation Details}
In the Tweet and User Meta\hyp{}Data Analysis section, we used the Face++ services to determine the number of faces and get demographics information in case of there is a single face in profile photo of the Twitter account.\footnote{Face++ \url{ https://www.faceplusplus.com}} In the location analysis section, we used Yandex geocoding services to convert geo\hyp{}coordinates and missing or incomplete location data into a standard format.\footnote{Yandex geocoding services  \url{https://tech.yandex.com/maps/geocoder}}

In the stance classification, sentiment detection, and topic discovery parts, we only used the tweets written in English.

In topic discovery section, we used the LDA algorithm \cite{Blei-03} provided by Gensim library \cite{Rehurek-2010}. In order to eliminate non\hyp{}influential tweets from topic discovery logic, we filtered out the tweets that are retweeted by other users for less than 10 times and containing less than 10 words. This criteria plays a role in eliminating non\hyp{}influential and short tweets from topic discovery algorithm. By using this dataset, we performed the following operations: preprocessing with the method of Gensim library, removing the stopwords, lemmatizing the words, and converting words to bigrams. Regarding the coherence score and the human judgment on the topics, we concluded that the LDA model achieves its best results with the following parameters: \textit{topic count=20, iteration count=500}.

At the beginning of our politician account analysis, we first divided the accounts that had more than ten thousand mentions into three categories: politicians, news channels, campaign/party accounts. We also analyzed the bot behavior of these accounts and found bot behavior in only two campaigns and one party account. (See Table \ref{table6}).

\begin{table*}
\captionsetup{justification=centering}
\caption{P1 \hyp{} Tweets posted between January and June 2016}
\scriptsize
\begin{center}
\begin{tabular}{ ll }
 \toprule
 Label & Top Representative Words\\
 \midrule 
1.Pro\hyp{}Leave side opinions about the election & vote,leave,britain,people,want,referendum,take,country,today,great,day,voter,future,win\\
2.Polarized opinions about the election & euref,voteleave,remain,leaveeu,strongerin,eureferendum,poll,eu,democracy,campaign,voteremain\\
3.Negative Impacts to Economy & big,london,lose,england,bad,pound,job,fall,bank,risk,rise,hit,blame,stock,drop\\
4.External & news,bbc,racist,everyone,consequence,german,donald\_trump,worry,french\\
5.Politics & cameron,political,tory,government,medium,believe,party,politician,second,elite\\
6.US & trump,support,thank,obama,wrong,explain,president,question,racism\\
7.Customs union & world,mean,post,market,economy,trade,deal,global,free,politic,union\\
8.Borders and economy & let,year,impact,economic,week,control,last,border,chance,cost,problem,cut,cause,collapse,problem\\
9.Handover of Cameron's PM position & david cameron,result,happen,pm,next,affect,become,resign\\
10.Pro\hyp{}Leave side opinions & british,exit,freedom,parliament,pay,independence,american,english,expect,ireland,turkey,reform\\
11.Brexit deal with Europe & europe,stop,end,migrant,merkel,nation,fight,war,destroy,negotiation,project,celebrate\\
12.Public policies & follow,nhs,immigration,botis,decision,ttip,farage,immigrant,issue,important,agree,promise,protect\\
13.Controversial opinions& time,talk,change,money,real,friend,life,state,save,possible,sovereignty,opportunity,divide,family\\
14.Indyref & scotland,stay,look,realdonaldtrump,happy,indyref,independent,scottish,xenophobia,public,message,refugee\\
15.Internal politics&business,fail,strong,article,continue,wake,minister,shock,juncker,threat,uncertainty,damage,army,queen\\
16.Divorce from EU & brit,brussel,borisjohnson,law,rule,davidcameron,labour,member,statement,mp,leader,corbyn,divorce\\
17. International & break,live,borishjohnson,history,france,germany,globalist,greece,spain,micheal\_gove,russia,international\\
18. Feelings and events & watch,euro,love,late,share,tonight,hate,attack,speech,begin,story,analysis,white,historic,police,spread,black\\
19. Internal politics & lie,fear,nigel\_farage,ukip,plan,bring,true,industry,chef,reveal,safe,worker,failure,angry,charge\\
20.Results of ref.&first,feel,morning,open,victory,claim,benefit,major,ready,regret\\
\bottomrule
\end{tabular}
\end{center}
\label{table8}
\end{table*}

\begin{table*}
\captionsetup{justification=centering}
\caption{P2 \hyp{} Tweets posted between June 2016 and February 2017}
\scriptsize
\begin{center}
\begin{tabular}{ ll }
 \toprule
 Label & Top Representative Words\\
 \midrule 
1.Pro\hyp{}Remain & people,theresa\_may,parliament,british,pm,stop,pm,democracy,remain,brexitshamble,negotiation,agree,majority\\
2.Pro\hyp{}Leave & ukip,leave,nigel\_farage,euref,great,lie,referendum,campaign,remain,voter,farage,poll,people\\
3.Personal opinions & vote,article,leaveeu,bill,remain,national,believe,trigger,people,labour,end,ignore,accuse,voting,june\\
4.Prospective plans  & plan,talk,today,theresamay,speech,ma,watch,idea,pm,andrealeadsom,need,time,listen,strategy,analysis\\
5.Proleave & want,britain,work,single\_market,stay,warn,european,minister,issue,country,access,state,brussel,brit,free\_movement\\
6.Economics & london,business,post,impact,move,bank,city,firm,job,huge,cost,financial,international,britain,economist,warn,company\\
7.Governance & government,rule,law,right,citizen,decision,court,challenge,power,new,post,irish,refuse,protect,high\_court\\
8.International politics & trump,world,future,europe,win,election,fascism,britain,global,trade\_agreement,leader,meet,speak,president,america\\
9.Immigration&immigration,bad,blame,report,policy,control,money,ireland,migrant,border,export,open,problem,pay,finance,migration\\
10.Crisis&year,economy,cost,stock,nhs,cut,pharma\_bank,por,region,due,uncertainty,investment,crisis,worker,funding,effect,investor\\
11.Europe&pound,fact,rise,fall,price,germany,sterling,euro,merkel,fear,increase,italy,value,home,drop,europe,passport\\
12.Trade deal&good,deal,trade,news,happen,free,bbc,next,britain,itv,discuss,post,negotiate,deliver,trade\_deal\\
13.Polarized & scotland,indyref,brexitcost,labour,debate,tory,independent,england,support,member,scottish,libdem,conservative,wale\\
14.Economic drawbacks & day,economic,find,question,research,damage,little,evidence,bbcnew,shock,consequence,economy,science,tomorrow,prepare\\
15.Negative feelings & lie,nonsense,join,interest,brexitbritain,french,history,europe,putin,interview,blair,official,russia,guardian,turkey,refugee\\
16.Internal politics & tory,politician,corbyn,pro,labour,spend,judge,cameron,nhs,houseoflord,tony\_blair,gina\_miller,unelect\\
17.Expect for change&politic,sign,many,change,hold,referendum,petition,democracy,sunderland,bregret,racist,war\\
18.Polarized & fight,remain\_yeseu,borisjohnson,wrong,hate,press,supporter,predict,racism,david\_cameron,lead,crash,voteleave,xenophobia\\
19.Negative impacts of ref. & lose,job,risk,eureferendum,late,expert,freedom,create,movement,brexiter,protest,understand,implication,sovereignty\\
20.New ref. request & become,tax,reason,remain,britain,tory,destroy,wish,marchforeurope,run,benefit,worry,ambassador,nobrexit\\
\bottomrule
\end{tabular}
\end{center}
\label{table9}
\end{table*}

\begin{table*}
\captionsetup{justification=centering}
\caption{P3 \hyp{} Tweets posted between February 2017 and November 2017}
\scriptsize
\begin{center}
\begin{tabular}{ll}
 \toprule
 Label & Top Representative Words\\
 \midrule 
1.Pro\hyp{}Remain & leave,remain,want,stopbrexit,people,lie,know,stop,country,support,campaign,remainer,fight,voter,leaver,help,politician\\
2.Labour & tory,labour,time,hard,party,conservative,corbyn,libdem,mp,stand,stop,sign,disaster,political,jeremycorbyn,policy,ukip\\
3.Future impacts & right,day,citizen,future,today,happen,live,eu,negotiation,debate,important,protect,reality,tomorrow\\
4.Pro\hyp{}Remain & vote,ge,referendum,may,election,call,poll,majority,win,result,remain,june,final,ukip,stopbrexit,mandate,voter,back,chance\\
5.Decisions about Ireland & new,ireland,report,read,post,government,law,late,border,parliament,today,paper,publish,cost,confirmirish,effect\\
6.Negotiations with EU & theresa\_may,talk,negotiation,pm,start,brussel,may,begin,letter,leader,gibraltar,negotiate,plan,deliver,hand\\
7.Request for a change& british,people,news,democracy,change,speak,believe,reason,government,union,decision,true,nation,briton,stupid\\
8.Economics & deal,trade,economy,britain,strong,economic,world,great,agree,head,minister,act,voting,damage,need,strategy,weak,self\_harm\\
9.Financial consequences & trump,nigel\_farage,fact,man,power,poor,rise,war,britain,people,side,inflation,brexiteer,history,european\_union,maga,rich\\
10.Impacts of leaving EU & good,europe,bad,look,join,france,germany,feel,idea,thing,britain,see,possible,news,exit,save,doctor,sad,influence,outcome\\
11. Economics & guardian,business,independent,ukip,stopbrexit,maydup,tax,cut,threat,economy,due,drop,stopbrexitnow,budget,git,grow,fund\\
12. Potential crisis & lose,warn,job,london,march,move,bank,risk,england,unite,staff,national,pound,company,britain,juncker,euro,parliament,big\\
13.Financial impacts& european,lord,food,house,speech,london,crisis,global,farmer,president,financial,discuss,impact,city,sector,fintech,school,dublin\\
14.Customs union& pay,keep,single\_market,stay,britain,bill,ukip\_leaveeu,rule,truth,ready,access,membership,eu,customs\_union,going\_backward\\
15.Scotland & scotland,scotref,ask,indyref,bbcnew,scottish,independence,answer,snp,scot,westminster,may,protest,government\\
16.Speculations & article,trigger,farage,trump,putin,thread,impact,tweet,russia,link,study,evidence,ukip,group,excellent,role,russian,author\\
17.Instability & fall,blame,stock,problem,negotiator,pharma\_bank,expect,irish,wale,borisjohnson,export,chiled,guarantee,family,uncertainty\\
18.News agencies& nhs,benefit,break,money,german,promise,block,love,billion,screw,racist,daily\_mail,timfarron,via\_reutersuk,sturgeon\\
19.Immigration & immigration,free,worker,control,freedom,fear,ukip,surprise,nhs,britain,movement,australia,sovereignty,migration,india\\
20.Social security & video,nhs,brexitshamble,nurse,reverse,boris,call,hammond,action,water,shock,murdoch,united,dream,brexitbritain,fox,ruin\\
\bottomrule
\end{tabular}
\end{center}
\label{table10}
\end{table*}

\end{document}